\documentclass[twoside,twocolumn,9pt]{article}
\usepackage{extsizes}
\usepackage[super,sort&compress,comma]{natbib} 
\usepackage[version=3]{mhchem}
\usepackage[left=1.5cm, right=1.5cm, top=1.785cm, bottom=2.0cm]{geometry}
\usepackage{balance}
\usepackage{times,mathptmx}
\usepackage{sectsty}
\usepackage{graphicx} 
\usepackage{lastpage}
\usepackage[format=plain,justification=justified,singlelinecheck=false,font={stretch=1.125,small,sf},labelfont=bf,labelsep=space]{caption}
\usepackage{float}
\usepackage{fancyhdr}
\usepackage{fnpos}
\usepackage[english]{babel}
\addto{\captionsenglish}{%
  
}
\usepackage{array}
\usepackage{droidsans}
\usepackage{charter}
\usepackage[T1]{fontenc}
\usepackage[usenames,dvipsnames]{xcolor}
\usepackage{setspace}
\usepackage[compact]{titlesec}
\usepackage[colorlinks,linkcolor=black,citecolor=black]{hyperref}
\usepackage{epstopdf}%This line makes .eps figures into .pdf - please comment out if not required.

\definecolor{cream}{RGB}{222,217,201}

\hypersetup{
    colorlinks,
    linkcolor= [rgb]{0.1804,0.1882,0.5725},
    citecolor=[rgb]{0.1804,0.1882,0.5725},
    urlcolor=[rgb]{0.1804,0.1882,0.5725}
}

\begin{document}

\pagestyle{fancy}
\thispagestyle{plain}
\fancypagestyle{plain}{

%%%HEADER%%%
%\fancyhead[C]{\includegraphics[width=18.5cm]{head_foot/header_bar}}
%\fancyhead[L]{\hspace{0cm}\vspace{1.5cm}\includegraphics[height=30pt]{head_foot/journal_name}}
%\fancyhead[R]{\hspace{0cm}\vspace{1.7cm}\includegraphics[height=55pt]{head_foot/RSC_LOGO_CMYK}}
\renewcommand{\headrulewidth}{0pt}
}
%%%END OF HEADER%%%

%%%PAGE SETUP - Please do not change any commands within this section%%%
\makeFNbottom
\makeatletter
\renewcommand\LARGE{\@setfontsize\LARGE{15pt}{17}}
\renewcommand\Large{\@setfontsize\Large{12pt}{14}}
\renewcommand\large{\@setfontsize\large{10pt}{12}}
\renewcommand\footnotesize{\@setfontsize\footnotesize{7pt}{10}}
\makeatother

\renewcommand{\thefootnote}{\fnsymbol{footnote}}
\renewcommand\footnoterule{\vspace*{1pt}% 
\color{cream}\hrule width 3.5in height 0.4pt \color{black}\vspace*{5pt}} 
\setcounter{secnumdepth}{5}

\makeatletter 
\renewcommand\@biblabel[1]{#1}            
\renewcommand\@makefntext[1]% 
{\noindent\makebox[0pt][r]{\@thefnmark\,}#1}
\makeatother 
\renewcommand{\figurename}{\small{Fig.}~}
\sectionfont{\sffamily\Large}
\subsectionfont{\normalsize}
\subsubsectionfont{\bf}
\setstretch{1.125} %In particular, please do not alter this line.
\setlength{\skip\footins}{0.8cm}
\setlength{\footnotesep}{0.25cm}
\setlength{\jot}{10pt}
\titlespacing*{\section}{0pt}{4pt}{4pt}
\titlespacing*{\subsection}{0pt}{15pt}{1pt}
%%%END OF PAGE SETUP%%%

%%%FOOTER%%%
\fancyfoot{}
%\fancyfoot[LO,RE]{\vspace{-7.1pt}\includegraphics[height=9pt]{head_foot/LF}}
%\fancyfoot[CO]{\vspace{-7.1pt}\hspace{13.2cm}\includegraphics{head_foot/RF}}
%\fancyfoot[CE]{\vspace{-7.2pt}\hspace{-14.2cm}\includegraphics{head_foot/RF}}
\fancyfoot[RO]{\footnotesize{\sffamily{1--\pageref{LastPage} ~\textbar  \hspace{2pt}\thepage}}}
\fancyfoot[LE]{\footnotesize{\sffamily{\thepage~\textbar\hspace{2pt} 1--\pageref{LastPage}}}}
\fancyhead{}
\renewcommand{\headrulewidth}{0pt} 
\renewcommand{\footrulewidth}{0pt}
\setlength{\arrayrulewidth}{1pt}
\setlength{\columnsep}{6.5mm}
\setlength\bibsep{1pt}
%%%END OF FOOTER%%%

%%%FIGURE SETUP - please do not change any commands within this section%%%
\makeatletter 
\newlength{\figrulesep} 
\setlength{\figrulesep}{0.5\textfloatsep} 

\newcommand{\topfigrule}{\vspace*{-1pt}% 
\noindent{\color{cream}\rule[-\figrulesep]{\columnwidth}{1.5pt}} }

\newcommand{\botfigrule}{\vspace*{-2pt}% 
\noindent{\color{cream}\rule[\figrulesep]{\columnwidth}{1.5pt}} }

\newcommand{\dblfigrule}{\vspace*{-1pt}% 
\noindent{\color{cream}\rule[-\figrulesep]{\textwidth}{1.5pt}} }

\makeatother
%%%END OF FIGURE SETUP%%%

%%%TITLE, AUTHORS AND ABSTRACT%%%
\twocolumn[
  \begin{@twocolumnfalse}
%\vspace{3cm}
\sffamily
\begin{tabular}{p{18.0cm} m{0cm}  }

  \noindent\LARGE{\textbf{Valley polarization of exciton-polaritons in monolayer WSe\textsubscript{2} in a tunable microcavity}} \\%Article title goes here instead of the text "This is the title"
\vspace{0.3cm} & \vspace{0.3cm}  \\

 \noindent\large{Mateusz~Kr\'ol,$^{\ast}$\textit{$^{,a}$} Katarzyna~Lekenta,\textit{$^{a}$} Rafa\l{}~Mirek,\textit{$^{a}$} Karolina~\L{}empicka,\textit{$^{a}$} Daniel~Stephan,\textit{$^{a}$} Karol~Nogajewski,\textit{$^{a,b}$} \newline Maciej~R.~Molas,\textit{$^{a,b}$} Adam~Babi\'{n}ski,\textit{$^{a}$} Marek~Potemski,\textit{$^{a,b}$} Jacek~Szczytko\textit{$^{a}$} and  Barbara~Pi\k{e}tka$^{\dagger}$\textit{$^{,a}$}} 
 \vspace{0.3cm}
 
 \normalsize{\textit{$^{a}$~Institute of Experimental Physics, Faculty of Physics, University of Warsaw, ul.~Pasteura 5, PL-02-093 Warsaw, Poland.}}
 
\normalsize{\textit{$^{b}$~Laboratoire National des Champs Magn\'etiques Intenses, CNRS-UGA-UPS-INSA-EMFL, Grenoble, France}

\small{$^{\ast}$ Mateusz.Krol@fuw.edu.pl}\newline 
\small{$^{\dagger}$ Barbara.Pietka@fuw.edu.pl}
}
 & \\%Author names go here instead of "Full name", etc.

\vspace{.25cm}

\noindent\normalsize{Monolayer transition metal dichalcogenides, known for exhibiting strong excitonic resonances, constitute a very interesting and versatile platform for investigation of light--matter interactions. In this work we report on a strong coupling regime between excitons in monolayer WSe\textsubscript{2} and photons confined in an open, voltage-tunable dielectric microcavity. The tunability of our system allows us to extend the exciton-polariton state over a wide energy range and, in particular, to bring the excitonic component of the lower polariton mode into  resonance with other excitonic transitions in monolayer WSe\textsubscript{2}. With selective excitation of spin-polarized exciton-polaritons we demonstrate the valley polarization when the polaritons from the lower branch come into resonance with a bright trion state in monolayer WSe\textsubscript{2} and valley depolarization when they are in resonance with a dark trion state.} 

\end{tabular}

 \end{@twocolumnfalse} \vspace{0.6cm}

  ]
%%%END OF TITLE, AUTHORS AND ABSTRACT%%%

%%%FONT SETUP - please do not change any commands within this section
\renewcommand*\rmdefault{bch}\normalfont\upshape
\rmfamily
\section*{}
\vspace{-1cm}

%%%MAIN TEXT%%%%
\section{Introduction}

Phenomena resulting from a strong coupling between light and matter lay foundations for cavity quantum electrodynamics and constitute a new building block for quantum information processing. In the coupled state, properties of excitations in the optically active material and photons confined in the cavity are significantly modified. In consequence, any of these particles represents an eigenmode of the system. Instead, new eigenmodes are formed known as exciton-polaritons.

A study of exciton-polaritons in  monolayer transition metal dichalcogenides (1L-TMDs) is nowadays of much interest because of highly desirable thermal, electronic, mechanical and optical properties these materials can provide for future  applications\cite{Wang_RevModPhys2018,Koperski_Nanophotonics2017, Schneider_2018}. 
Monolayer TMDs are characterized by substantial oscillator strength of absorption resonances and excitonic binding energies as large as a few hundreds of meV's\cite{Chernikov_PRL2014,He_PRL2014}, which allows for realization of a strongly coupled light-matter system even at room temperature\cite{Lundt_NatComm2016}. %Moreover, the exciton saturation density in 1L-TMDs is much higher than in conventional semiconductor quantum wells. 

The strong light-matter coupling in a system comprising TMD material was realized for the first time in a dielectric cavity enclosing a monolayer MoS\textsubscript{2} flake \cite{Liu_NatPhoton2014}.  Shortly after, a similar effect was observed for a zero-dimensional cavity mode coupled to excitonic transitions in 1L MoSe\textsubscript{2}, when the cavity resonance was tuned by changing the cavity width \cite{Dufferwiel_NatComm2015}. Few experimental works followed that reports and the strong light-matter coupling phenomena in various types of microcavities containing MoSe\textsubscript{2} \cite{Sidler_NatPhys2016,Dufferwiel_NatPhoton2017,Lundt_2DMater2017}, WS\textsubscript{2} \cite{Flatten_SciRep2016,Lundt_PRB2017,Wang_OptExpress2016,Sun_NatPhoton2017,Chen_NatPhoton2017}  and WSe\textsubscript{2} \cite{Lundt_NatComm2016} were observed. TMDs were also used to obtain hybrid exciton--polaritons\cite{Flatten_NatComm2017,Wurdack_NatComm2017,Waldherr_NatCom2018}, in which mixing of exciton states from two different materials occurs through the cavity photon mode. The regime of a strong coupling between light and  excitons in WSe\textsubscript{2} flakes was also demonstrated  with plasmonic nanoparticles placed on metallic mirrors\cite{Kleemann_NatComm2017} and in a photonic crystal\cite{Zhang_NatComm2018}. 

Monolayer TMDs are direct-bandgap semiconductors with the lowest-energy excitonic transitions taking place in two $K$ an $-K$ valleys in the Brillouin zone. This additional valley degree of freedom can be used to store and carry information, a concept referred to as valleytronics\cite{Rycerz_NatPhys2007,Schaibley_NatRevMater2016}. Thanks to a strong spin-orbit interaction, which locks the particle spin to the valley pseudospin\cite{Xu_NatPhys2014}, addressing carriers to a specific valley in 1L-TMDs is possible via  an optical excitation of properly chosen helicity. Indeed, optical selection rules allow the $\sigma^+$ ($\sigma^-$) circularly polarized light to interact with carriers in the $K$ ($-K$) valley\cite{Xiao_PRL2012,Cao_NatComm2012}. A valley polarization created  this way is not only retained in the photoluminescence\cite{Mak_NatNanotech2012,Manca_NatComm2017} but also enables the formation of an intervalley coherent state by a linearly polarized excitation  \cite{Jones_NatNanotech2013,Wang_PRL2015}.

%In Ref.~\citenum{Bleu_PRB2016} Bleu \textit{et al.} predicted the optical valley Hall effect for TMDs exciton-polaritons. Due to the SOC, excitons with different polarizations placed in an optical cavity will exhibit a coherent precession of their pseudo-spin induced by the wave-vector dependent effective magnetic field.

% The WSe\textsubscript{2} is a particular type of two-dimensional crystal where the valley selective scattering \cite{Jones_NatNanotech2013,Wang_PRL2015,Manca_NatComm2017} plays a special role.

%The mutual coupling of spin and valley degrees of freedom is strong and leads to suppressed spin relaxation\cite{Xiao_PRL2012, Bleu_PRB2016}. 
The presence of a microcavity in a strongly coupled light-monolayer-TMD system  affects the intervalley relaxation processes in the monolayer through the reduction of intravalley scattering by disorder\cite{Bleu_PRB2016}, and hence results in long valley pseudospin relaxation times\cite{Dufferwiel_NatPhoton2017}. %, small effective mass and high in-plane momentum of exciton-polaritons. 
Additionally, since the intervalley scattering occurs only through the excitonic component of exciton-polariton, a higher valley polarization can be observed at room temperature for monolayer embedded  in a microcavity than for a bare one\cite{Chen_NatPhoton2017,Lundt_PRB2017}. Therefore, the spin-preservation and spin-mixing processes in 1L-TMD-based exciton-polaritons attract significant attention\cite{Sun_NatPhoton2017} and are important for possible applications in spin-based optoelectronic devices.

\begin{figure}%[h]
\centering
\includegraphics[width=.45\textwidth]{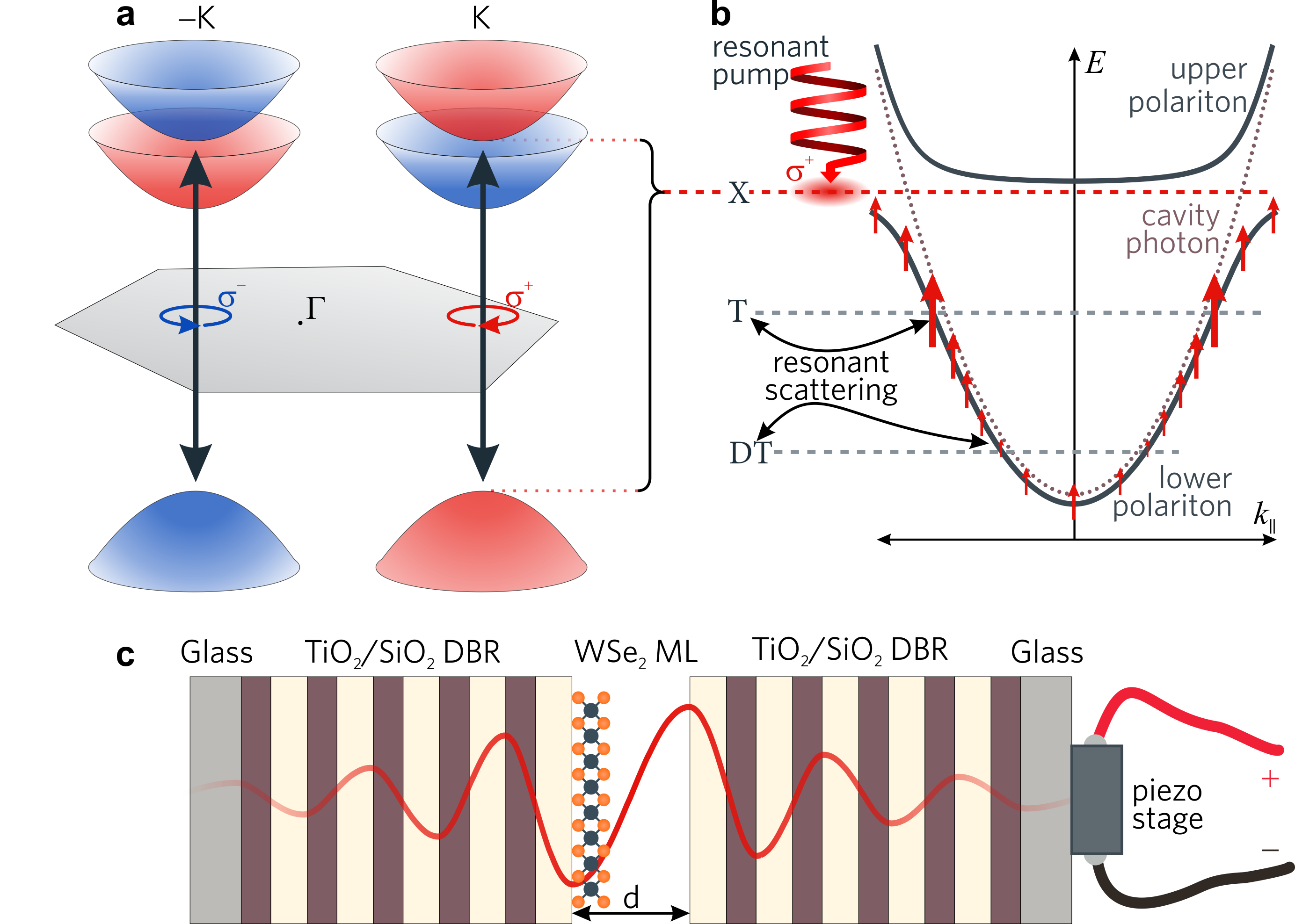}
\caption{\textbf{a} Scheme of the valley-selective excitonic transitions in a WSe\textsubscript{2} monolayer. \textbf{b} Coupling of excitonic transitions to a cavity mode. The spin-preserving carrier relaxation along the lower polariton branch is marked by red arrows. \textbf{c} Scheme of the open, tunable cavity used in the study described in this paper. The structure is designed to maximize the electric field distribution (schematically illustrated by the red wavy line) at the position of the monolayer.}
\label{im:Fig0}
\end{figure}

In this work we demonstrate a strongly coupled system of neutral excitons ($X$) in monolayer WSe\textsubscript{2} and photons confined in a tunable dielectric microcavity. By probing the photonic in-plane momentum we show the anti-crossing of the photonic mode with the $X$ resonance of WSe\textsubscript{2}, indicative of creation of exciton-polaritons. Moreover, thanks to  special construction of our planar cavity we are able to tune the cavity width by external voltage and to consequently change the photonic mode energy by more than 80\,meV, therefore changing the coupling conditions. Due to that, the polariton mode, containing a fraction of the neutral exciton's wave function, can be brought into resonance with other excitons in 1L WSe\textsubscript{2} such as a trion ($T$) and a dark trion ($DT$). A resonant injection of spin-polarized neutral excitons into the monolayer allows us to observe a resonant scattering  between the lower polariton mode and the $T$ and $DT$ states. We demonstrate that the spin valley polarization is enhanced when the polariton energy coincides with the $T$ and decreases when the same happens with the $DT$ resonance.

The valley spin polarization in 1L\,WSe\textsubscript{2}  is illustrated in Fig.~\ref{im:Fig0}a. Spin polarized excitons are formed in different  valleys in the momentum space: $K$ and $-K$. In a microcavity, those two types of excitons couple separately to photons of corresponding polarization, i.e. the spin-up (spin-down) excitons couple to the $\sigma^+$ ($\sigma^-$) photons. The strong coupling condition for spin-up excitons (red dashed line) and the $\sigma^+$ cavity photonic mode (violet dotted curve) is illustrated in Fig.~\ref{im:Fig0}b. The resulting exciton-polariton dispersion branches (black solid curves) can extend over a wide energy range and intersect with the $T$ and $DT$ energies (gray dashed lines). %Polaritons at given momentum can therefore scatter with trion-related resonances ($T$ and $DT$). which leads to increased and is reduced significantly, respectively, polarization degree in polariton mode.

\section{Results and discussion}
\subsection{Tunable cavity}

\begin{figure}%[h]
\centering
\includegraphics[width=.5\textwidth]{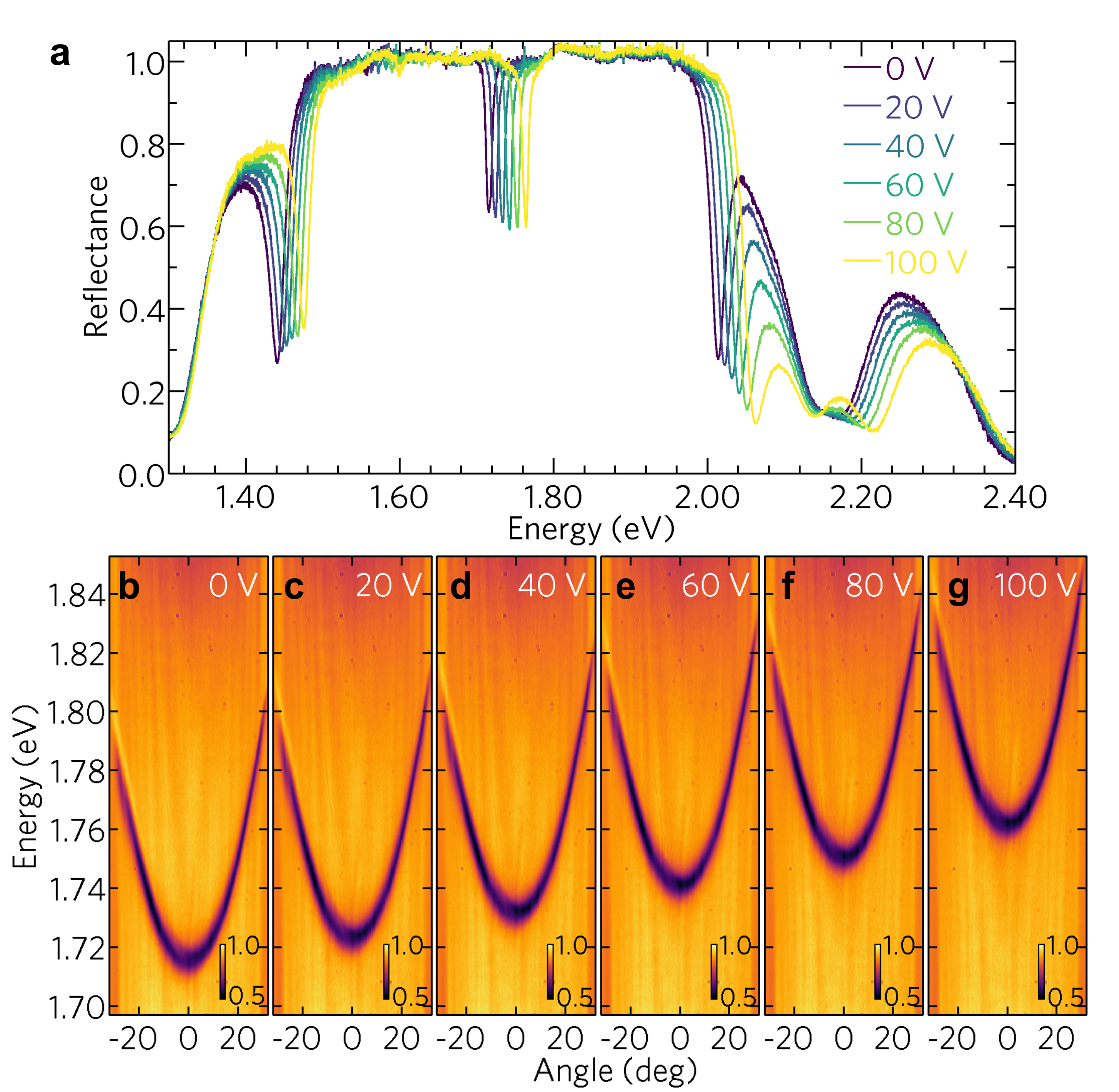}
\caption{Reflectance from an empty cavity at 5\,K. \textbf{a} Reflectance spectra at normal incidence for different voltages applied to the piezoelectric chip, \textbf{b}--\textbf{g} corresponding angle-resolved reflectance spectra for the cavity mode.}
\label{im:Fig1}
\end{figure}

A schematic illustration of the  investigated  structure with a tunable dielectric cavity is presented in Fig.~\ref{im:Fig0}c. The full cavity consists of two identical distributed Bragg reflectors (DBRs) centered at 1.714\,eV and composed of 5 pairs of TiO\textsubscript{2}/SiO\textsubscript{2} layers grown on a transparent glass substrate. Each DBR is terminated with an SiO\textsubscript{2} layer to ensure the maximum electric field's amplitude at the position of a WSe\textsubscript{2} monolayer deposited on top of a bottom mirror. The DBR with the monolayer  is placed on a piezoelectric actuator, which allows to control its separation from the second DBR by applying external voltage.

The operation of the empty cavity (no WSe\textsubscript{2} monolayer inside) at 5\,K is presented in Fig.~\ref{im:Fig1}. In the reflectance spectra shown in Fig.~\ref{im:Fig1}a, the cavity mode is visible as a narrow dip in the photonic stop-band  around 1.754\,eV.
As can be seen, with the external voltage applied to the chip, the position of the cavity mode shifts towards higher energies. Importantly, the tuning of the cavity has no significant impact on the cavity's quality factor, which at normal incidence is close to 160. In the angle-resolved experiment, realized by means of Fourier-plane imaging of a high-numerical-aperture (${\rm NA} = 0.55$) microscope objective, the reflectance spectra display a typical, parabolic-like energy vs observation angle dependence shown in Fig.\,\ref{im:Fig1}b--g. A blueshift of the cavity mode with the voltage polarizing the piezoelectric chip is clearly visible.

\subsection{Monolayer WSe\textsubscript{2} on a DBR}

\begin{figure}%[h]
\centering
\includegraphics[width=.5\textwidth]{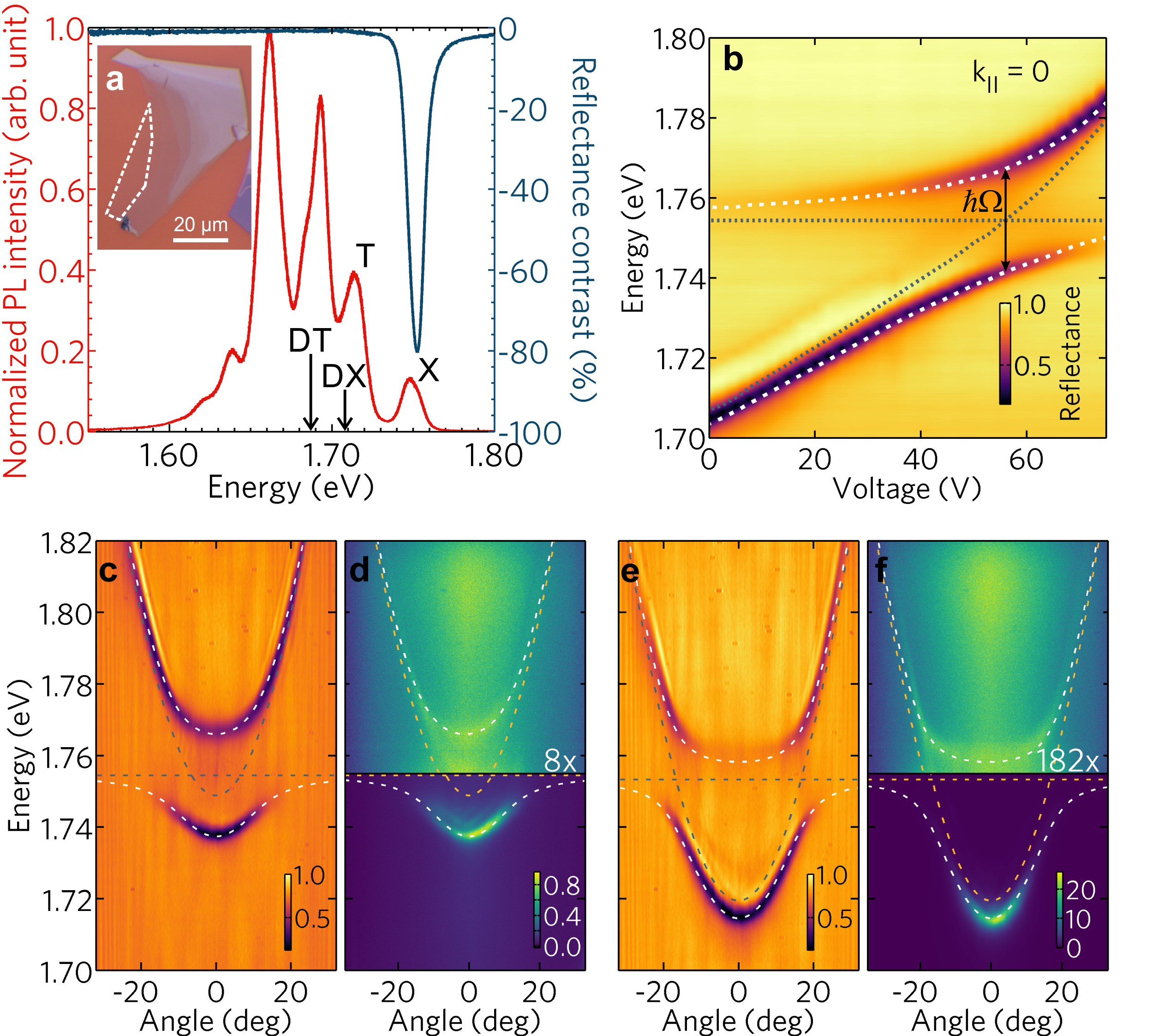}
\caption{\textbf{a} Photoluminescence and reflectance contrast spectra of the WSe\textsubscript{2} monolayer deposited on top of the dielectric mirror measured at 5\,K. The inset presents an optical microscope image of the flake with the single-layer periphery marked by the dashed line. \textbf{b} Reflectance at normal incidence for different voltages applied to the piezoelectric actuator. Reflectance and photoluminescence spectra for \textbf{c},\textbf{d} $-$5.7\,meV  and \textbf{e},\textbf{f} $-$33.9\,meV exciton--photon detuning.}
\label{im:Fig2}
\end{figure}

WSe\textsubscript{2} flakes containing monolayer terraces were exfoliated from a bulk crystal purchased from HQ Graphene and transfered on top of the DBR using an all-dry polydimethylsiloxane-based stamping method. An optical microscope image of one of such flakes is shown in the inset of Fig.~\ref{im:Fig2}a, with a white dashed line marking the boundary of the monolayer part. Drawn in red and blue in Fig.~\ref{im:Fig2}a are photoluminescence (PL) and reflectance contrast (RC) spectra measured at 5\,K. The PL spectrum displays several emission lines with a characteristic pattern already discussed in a number of previous works\citep{Arora_Nanoscale2015,Smolenski_PRX2016,Koperski_2018}. In accordance with those
reports, the highest energy emission peak ($X$) is related to a neutral exciton, the following one ($T$) is attributed to a trion complex and a series of lower-energy lines are assigned to the so-called localized exciton states. The energy positions of a dark neutral exciton ($DX$) and dark trion ($DT$), invoked in the following parts of the present paper, are also indicated although none of them is optically-active. In contrast to the PL results, the RC spectrum  shows just a single absorption peak at 1.754\,eV corresponding to the  $X$ resonance.

\subsection{Strong coupling regime}

The strong light--matter coupling in the system can be observed experimentally as an avoided crossing between the excitonic and photonic modes, as their energy difference at $k_{\|} = 0$ (often referred to as detuning~$\delta$) approaches  zero. The effect of tuning the cavity mode across the excitonic transition on the reflectance spectra measured at  normal incidence is presented in Fig.~\ref{im:Fig2}b. It reveals the behaviour characteristic of the strong coupling regime: when the cavity mode approaches the excitonic transition, two branches exhibiting a well-developed anti-crossing become  clearly distinguishable. 

The energy difference between the coupled modes at the minimum separation energy gives the exciton-photon coupling strength, described by the vacuum-field Rabi energy, $\hbar\Omega$. By fitting a two-mode polariton model \cite{Weisbuch_PRL1992} with a constant exciton energy $E_{X} = 1.754$\,meV to the experimental data presented in Fig.\,\ref{im:Fig2}b we get the  coupling strength $\hbar\Omega = 26.5$\,meV. Worth noting is that Fig.\,\ref{im:Fig2}b also provides a straightforward way of converting the voltage applied to the piezo-actuator to detuning between the excitonic resonance and the cavity mode.

The anti-crossing can be directly seen also in the angle-resolved reflectance spectra as shown in Fig.~\ref{im:Fig2}c and \ref{im:Fig2}\,e for two different detunings. In both figures two well-separated polariton branches (the lower and upper one) are clearly visible. Fitting polariton dispersion relations to these modes in both cases gives the coupling strength $\hbar\Omega = 28$\,meV, which stays in good agreement with the value obtained by analysing the data presented in Fig.\,\ref{im:Fig2}b. Positions of the modes in the reflectance measurements coincide with the emission lines observed in photoluminescence maps, as illustrated in Fig.~\ref{im:Fig2}d\,and \ref{im:Fig2}\,f. In comparison with the angle-resolved reflectance spectra, the intensities of the PL spectra significantly vary between the two polariton branches. The strongest emission occurs at the  bottom of the lower polariton branch. This is the lowest energy state available to polaritons, which is the most occupied due to  relaxation processes taking place along the branch. A weak emission from the upper polariton branch can also be  observed, but depending on the detuning, its intensity falls behind that from the lower branch by one to two orders of magnitude.

\subsection{Valley polarization}

\begin{figure}[t]
\centering
\includegraphics[width=.5\textwidth]{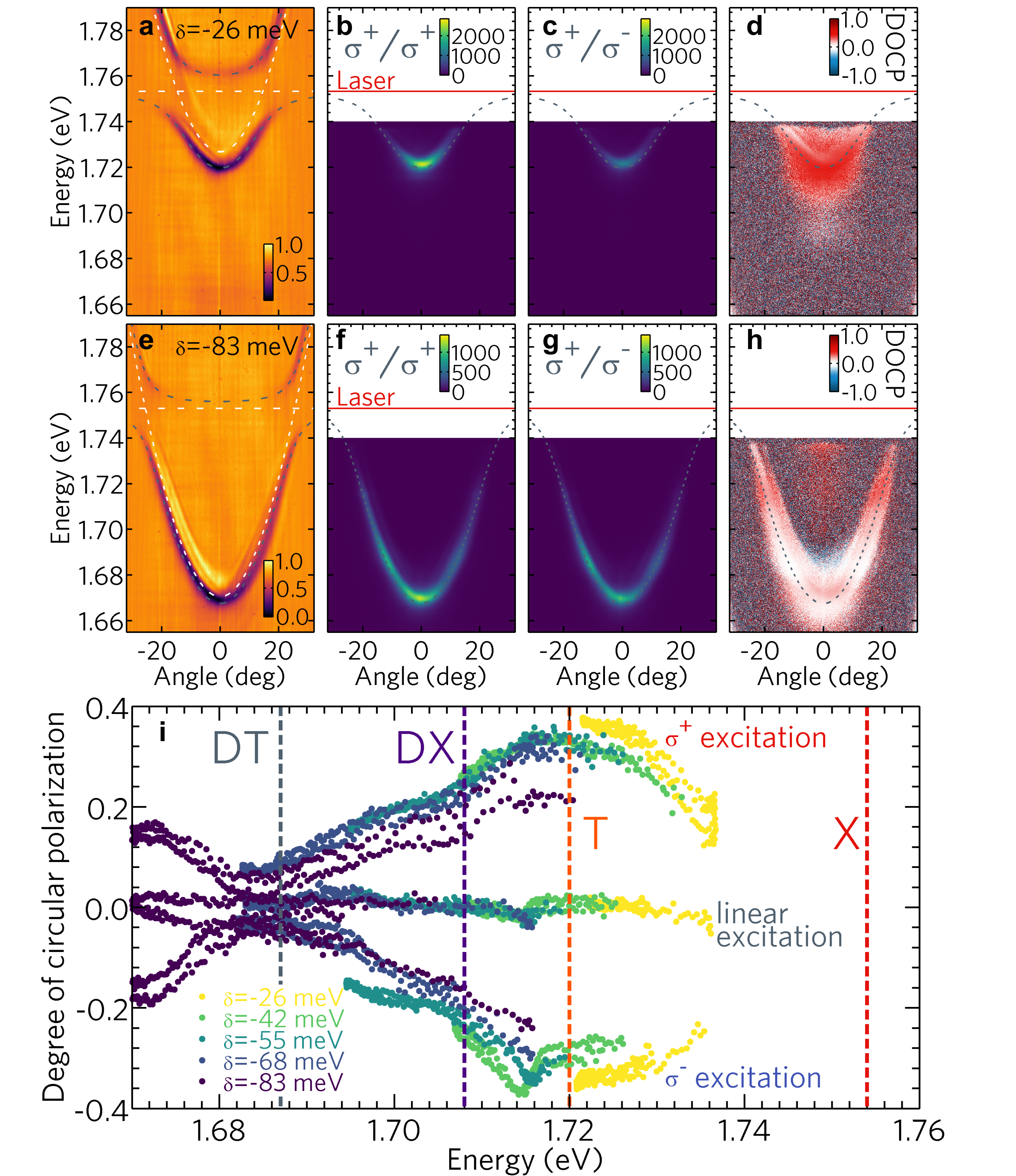}
\caption{Angle-resolved: \textbf{a,e} reflectance, photoluminescence excited in $\sigma^+$  detected in \textbf{b,f} $\sigma^+$ and \textbf{c,g} $\sigma^-$ polarization with resulting \textbf{d,h} degree of circular polarization (DOCP) at  exciton--photon detunings of \textbf{a--d} $-$26\,meV or \textbf{e--h} $-$83\,meV. \textbf{i} DOCP dependence on emission energy for different detunings and $\sigma^+$, $\sigma^-$  and linear excitation. Energy positions of the trion $T$, dark exciton $DX$ and dark trion $DT$ in 1L WSe\textsubscript{2} are based on Ref\,\cite{Zhang_NatNanotech2017}.}
\label{im:Fig3}
\end{figure}

The exciton-polariton spin structure is analogous to that of an exciton. Projections of the exciton total angular momentum on the quantization axis perpendicular to the plane of DBRs and the monolayer can be equal to $\pm1$ or $\pm2$. The $\pm$2 excitons remain dark and do not couple to cavity photons. The optically-active excitons, $\pm$1, couple respectively to $\sigma^\pm$ circularly polarized light, forming exciton-polaritons of two opposite spins. The relaxation of exciton-polaritons close to their ground state is modified with respect to that of bare excitons due to the shape of their dispersion curve \cite{Kavokin_PRL2004}. However, the main relaxation mechanisms, such as the scattering with acoustic phonons, free electrons and polariton-polariton collisions are still the most efficient ones. It was shown for exciton-polaritons in semiconductor microcavities, that the polariton relaxation mechanisms acting along the lower polariton branch conserve the total spin \cite{Kavokin_PRL2004, Shelykh_PRB2004}. In an exciton-polariton system comprising a material with strong valley selectivity, like WSe\textsubscript{2}, the energy relaxation processes  should  also preserve the spin and as such should be restricted to only  one valley. This spin-conserving scenario is valid unless the polariton comes into resonance with another state it can scatter to. Under such resonant conditions the spin flip processes resulting from collisions of the excitonic part of an exciton-polariton with structure imperfections, impurities, phonons or other types of excitations in the system may occur \cite{Kavokin_PRL2004} leading to depolarization of emitted light. 

In order to investigate the exciton-polariton's spin conservation related to selective valley polarization, we performed a series of PL measurements under resonant excitation. The energy of circularly polarized excitation light generated by a tunable single mode Ti:sapphire laser was set to match the neutral exciton's ($X$) energy in the WSe\textsubscript{2} monolayer, estimated from the angle-resolved reflectance maps presented in Figs.~\ref{im:Fig3}a and \ref{im:Fig3}\,e. The emission  signal was collected in both circular polarizations which allowed for calculation of the degree of circular polarization (DOCP) defined as $\textrm{DOCP} = \frac{I_{\sigma^+}-I_{\sigma^-}}{I_{\sigma^+}+I_{\sigma^-}}$, and shown in Fig.~\ref{im:Fig3}d,h for two different detunings. According to the optical selection rules, a circularly polarized and resonant excitation beam excites the carriers only in one of the two valleys in the reciprocal space. %Due to a high momentum separation between the $K$ and $-K$ points in  the Brillouin zone, the intervalley carrier scattering is typically strongly suppressed. 
The $\sigma^+$ ($\sigma^-$) circular polarization of the optical excitation is preserved to some extent in the spin of photo-created excitons and results in a positive (negative) sign of DOCP of light emitted after their recombination, as can be seen in Figs.\,\ref{im:Fig3}d and \ref{im:Fig3}h. 

DOCP of the emitted light calculated from the PL data recorded at four different exciton--photon detunings is presented as a function of collected photon energy in Fig.~\ref{im:Fig3}i. For the $\sigma^+$ polarized pump the highest value of DOCP, equal to about 0.35, occurs at 1.72\,eV, and the minimum is observed at around 1.687\,eV (accordingly, the orange and gray dashed lines in Fig.~\ref{im:Fig3}i). Based on the energy distance to the neutral exciton line, those two energies correspond to the bright ($T$) and dark ($DT$) trion, as  revealed by optical experiments carried out in magnetic field applied in the Voigt configuration\cite{Zhang_NatNanotech2017}. Analogus increase of DOCP at the energy of trion in a similar system incorporating MoSe\textsubscript{2} monolayers inside a tunable cavity has  already been reported in Ref.\,\cite{Dufferwiel_NatPhoton2017}. Such an observation is also consistent with a time-resolved photouminescence study of WSe\textsubscript{2} monolayers, showing  much slower valley depolarization time for charged excitons\cite{Wang_PRB2014}, especially for intervalley trions\cite{Singh_PRL2016}, than for the neutral exciton. As already suggested, the decrease in the valley polarization observed at about 1.687\,eV can be in our opinion attributed to resonant scattering between the photon-coupled polarized bright excitons and the dark trion state. A decrease of the emission intensity expected for such a scattering mechanism is shown in Figs~S1--S3\dag. It should be noted here, that different experiments\cite{Wang_PRL2017,Zhou_NatureNanotech2017}, especially in magnetic field \cite{Zhang_NatNanotech2017,Molas_2DMater2017}, reveal also the existence of a dark exciton transition ($DX$) 40--47\,meV below the neutral exciton ($X$), which seems to have no or only a little impact on our polarization measurements. An important remark is, that the DOCP behaviour observed for the $\sigma^-$ excitation is identical to that for the $\sigma^+$ pump, but as expected has the opposite sign (see Fig.~\ref{im:Fig3}i).

A linearly polarized excitation beam, which represents a coherent superposition of both circular polarizations, populates with carriers both valleys equally, which leads to a PL signal which is not spin-polarized. 

Let us also comment on the importance of the TE-TM splitting on the relaxation mechanisms. In planar microcavities the TE-TM splitting of the cavity mode acts as an effective magnetic field in the cavity plane, whose magnitude increases with the observation angle\cite{Kavokin_PRL2005}. Such a splitting can lead to a slightly increased valley depolarization for higher-wavevector polaritons.  Based on transfer matrix calculations we estimated the  TE-TM splitting to reach in our case up to 2\,meV, which is significantly lower than the polariton linewidth equal to about 7\,meV, what suggests that the magnitude of the effective magnetic field caused by this effect is relatively low.

\section{Conclusions}

In summary, we have demonstrated a strong coupling between neutral excitons in monolayer WSe\textsubscript{2} and photons confined in an open dielectric cavity. A possibility to tune the cavity resonance, with access to dispersion relations via angle-resolved experiments, allowed us to create the exciton-polariton  modes that can extend over a wide energy range and, in particular,  cross with other excitonic resonances in the WSe\textsubscript{2} monolayer. We have shown that selective excitation of spin-polarized exciton-polaritons leads to the observation of enhanced valley polarization for the resonance of the lower polariton branch with a bright trion and valley depolarization caused by resonant scattering of polaritons to a dark trion state. Our results demonstrate an important role of spin-selective mechanisms in the resonant scattering of excitonic complexes in monolayer transition metal dichalcogenides.% In particular, the enhancement of the spin preservation can be used in a ... and in optical manipulation ... %Di Xiao  integration of valleytronics and spintronics in multivalley materials with strong spin-orbit coupling
% Bleu where the information is stored in the valley degree of freedom of carriers, which can offer a better protection against relaxation
% ref. opto-valleytronics: z Bleu: K. F. Mak, K. He, J. Shan, and T. F. Heinz, Nat. Nanotechnology 7, 494 (2012). [24] G. Sallen, L. Bouet, X. Marie, G. Wang, C. R. Zhu, W. P. Han, Y. Lu, P. H. Tan, T. Amand, B. L. Liu, and B. Urbaszek, Phys. Rev. B 86, 081301 (2012). [25] G. Kioseoglou, A. T. Hanbicki, M. Currie, A. L. Friedman, D. Gunlycke, and B. T. Jonker, Appl. Phys. Lett. 101, 221907 (2012). [26] D. Lagarde, L. Bouet, X. Marie, C. R. Zhu, B. L. Liu, T. Amand, P. H. Tan, and B. Urbaszek, Phys. Rev. Lett. 112, 047401 (2014).

\section*{Conflicts of interest}
There are no conflicts to declare.

\section*{Acknowledgements}

Scientific work co-financed from the Ministry of Higher Education budget for education as a research project "Diamentowy Grant" 0109/DIA/2015/44. This work was supported by the National Science Centre grant UMO-2013/10/M/ST3/00791 and EU Graphene Flagship project (ID: 785219). B.\,P. acknowledges Ambassade de France en Pologne for the research stay project. The support from the Foundation for Polish Science through the project "The atomically thin semiconductors for future optoelectronics" carried out within the TEAM programme and cofinanced by the European Union under the European Regional Development Fund is also acknowledged.

%%%REFERENCES%%%
%\bibliography{bib_abrev} 
%\bibliographystyle{rsc} %rsc the RSC's .bst file

\providecommand*{\mcitethebibliography}{\thebibliography}
\csname @ifundefined\endcsname{endmcitethebibliography}
{\let\endmcitethebibliography\endthebibliography}{}

\end{document}

\pagestyle{fancy}
\thispagestyle{plain}
\fancypagestyle{plain}{

%%%HEADER%%%
%\fancyhead[C]{\includegraphics[width=18.5cm]{head_foot/header_bar}}
%\fancyhead[L]{\hspace{0cm}\vspace{1.5cm}\includegraphics[height=30pt]{head_foot/journal_name}}
%\fancyhead[R]{\hspace{0cm}\vspace{1.7cm}\includegraphics[height=55pt]{head_foot/RSC_LOGO_CMYK}}
\renewcommand{\headrulewidth}{0pt}
}
%%%END OF HEADER%%%

%%%PAGE SETUP - Please do not change any commands within this section%%%
\makeFNbottom
\makeatletter
\renewcommand\LARGE{\@setfontsize\LARGE{15pt}{17}}
\renewcommand\Large{\@setfontsize\Large{12pt}{14}}
\renewcommand\large{\@setfontsize\large{10pt}{12}}
\renewcommand\footnotesize{\@setfontsize\footnotesize{7pt}{10}}
\makeatother

\renewcommand{\thefootnote}{\fnsymbol{footnote}}
\renewcommand\footnoterule{\vspace*{1pt}% 
\color{cream}\hrule width 3.5in height 0.4pt \color{black}\vspace*{5pt}} 
\setcounter{secnumdepth}{5}

\makeatletter 
\renewcommand\@biblabel[1]{#1}            
\renewcommand\@makefntext[1]% 
{\noindent\makebox[0pt][r]{\@thefnmark\,}#1}
\makeatother 
\renewcommand{\figurename}{\small{Fig.}~}
\sectionfont{\sffamily\Large}
\subsectionfont{\normalsize}
\subsubsectionfont{\bf}
\setstretch{1.125} %In particular, please do not alter this line.
\setlength{\skip\footins}{0.8cm}
\setlength{\footnotesep}{0.25cm}
\setlength{\jot}{10pt}
\titlespacing*{\section}{0pt}{4pt}{4pt}
\titlespacing*{\subsection}{0pt}{15pt}{1pt}
%%%END OF PAGE SETUP%%%

%%%FOOTER%%%
\fancyfoot{}
%\fancyfoot[LO,RE]{\vspace{-7.1pt}\includegraphics[height=9pt]{head_foot/LF}}
%\fancyfoot[CO]{\vspace{-7.1pt}\hspace{13.2cm}\includegraphics{head_foot/RF}}
%\fancyfoot[CE]{\vspace{-7.2pt}\hspace{-14.2cm}\includegraphics{head_foot/RF}}
\fancyfoot[RO]{\footnotesize{\sffamily{1--\pageref{LastPage} ~\textbar  \hspace{2pt}\thepage}}}
\fancyfoot[LE]{\footnotesize{\sffamily{\thepage~\textbar\hspace{2pt} 1--\pageref{LastPage}}}}
\fancyhead{}
\renewcommand{\headrulewidth}{0pt} 
\renewcommand{\footrulewidth}{0pt}
\setlength{\arrayrulewidth}{1pt}
\setlength{\columnsep}{6.5mm}
\setlength\bibsep{1pt}
%%%END OF FOOTER%%%

%%%FIGURE SETUP - please do not change any commands within this section%%%
\makeatletter 
\newlength{\figrulesep} 
\setlength{\figrulesep}{0.5\textfloatsep} 

\newcommand{\topfigrule}{\vspace*{-1pt}% 
\noindent{\color{cream}\rule[-\figrulesep]{\columnwidth}{1.5pt}} }

\newcommand{\botfigrule}{\vspace*{-2pt}% 
\noindent{\color{cream}\rule[\figrulesep]{\columnwidth}{1.5pt}} }

\newcommand{\dblfigrule}{\vspace*{-1pt}% 
\noindent{\color{cream}\rule[-\figrulesep]{\textwidth}{1.5pt}} }

\makeatother
%%%END OF FIGURE SETUP%%%

%%%TITLE, AUTHORS AND ABSTRACT%%%
\twocolumn[
  \begin{@twocolumnfalse}
%\vspace{3cm}
\sffamily
\begin{tabular}{p{18.0cm} m{0cm}  }

  \noindent\LARGE{\textbf{Supplementary information for Valley polarization of exciton-polaritons in monolayer WSe\textsubscript{2} in a tunable microcavity}} \\%Article title goes here instead of the text "This is the title"
\vspace{0.3cm} & \vspace{0.3cm}  \\

 \noindent\large{Mateusz~Kr\'ol,$^{\ast}$\textit{$^{,a}$} Katarzyna~Lekenta,\textit{$^{a}$} Rafa\l{}~Mirek,\textit{$^{a}$} Karolina~\L{}empicka,\textit{$^{a}$} Daniel~Stephan,\textit{$^{a}$} Karol~Nogajewski,\textit{$^{a,b}$} \newline Maciej~R.~Molas,\textit{$^{a,b}$} Adam~Babi\'{n}ski,\textit{$^{a}$} Marek~Potemski,\textit{$^{a,b}$} Jacek~Szczytko\textit{$^{a}$} and  Barbara~Pi\k{e}tka$^{\dagger}$\textit{$^{,a}$}} 
 \vspace{0.3cm}
 
 \normalsize{\textit{$^{a}$~Institute of Experimental Physics, Faculty of Physics, University of Warsaw, ul.~Pasteura 5, PL-02-093 Warsaw, Poland.}}
 
\normalsize{\textit{$^{b}$~Laboratoire National des Champs Magn\'etiques Intenses, CNRS-UGA-UPS-INSA-EMFL, Grenoble, France}

\small{$^{\ast}$ Mateusz.Krol@fuw.edu.pl}\newline
\small{$^{\dagger}$ Barbara.Pietka@fuw.edu.pl}
}
 & \\%Author names go here instead of "Full name", etc.

\end{tabular}

 \end{@twocolumnfalse} \vspace{0.6cm}

  ]
%%%END OF TITLE, AUTHORS AND ABSTRACT%%%

%%%FONT SETUP - please do not change any commands within this section
\renewcommand*\rmdefault{bch}\normalfont\upshape
\rmfamily
\section*{}
\vspace{-1cm}

\renewcommand{\thefigure}{S\arabic{figure}}

%%%END OF FOOTNOTES%%%

\section*{Emission suppression for the resonance with the dark trion state}

\begin{figure}[H]
\centering
\includegraphics[width=.41\textwidth]{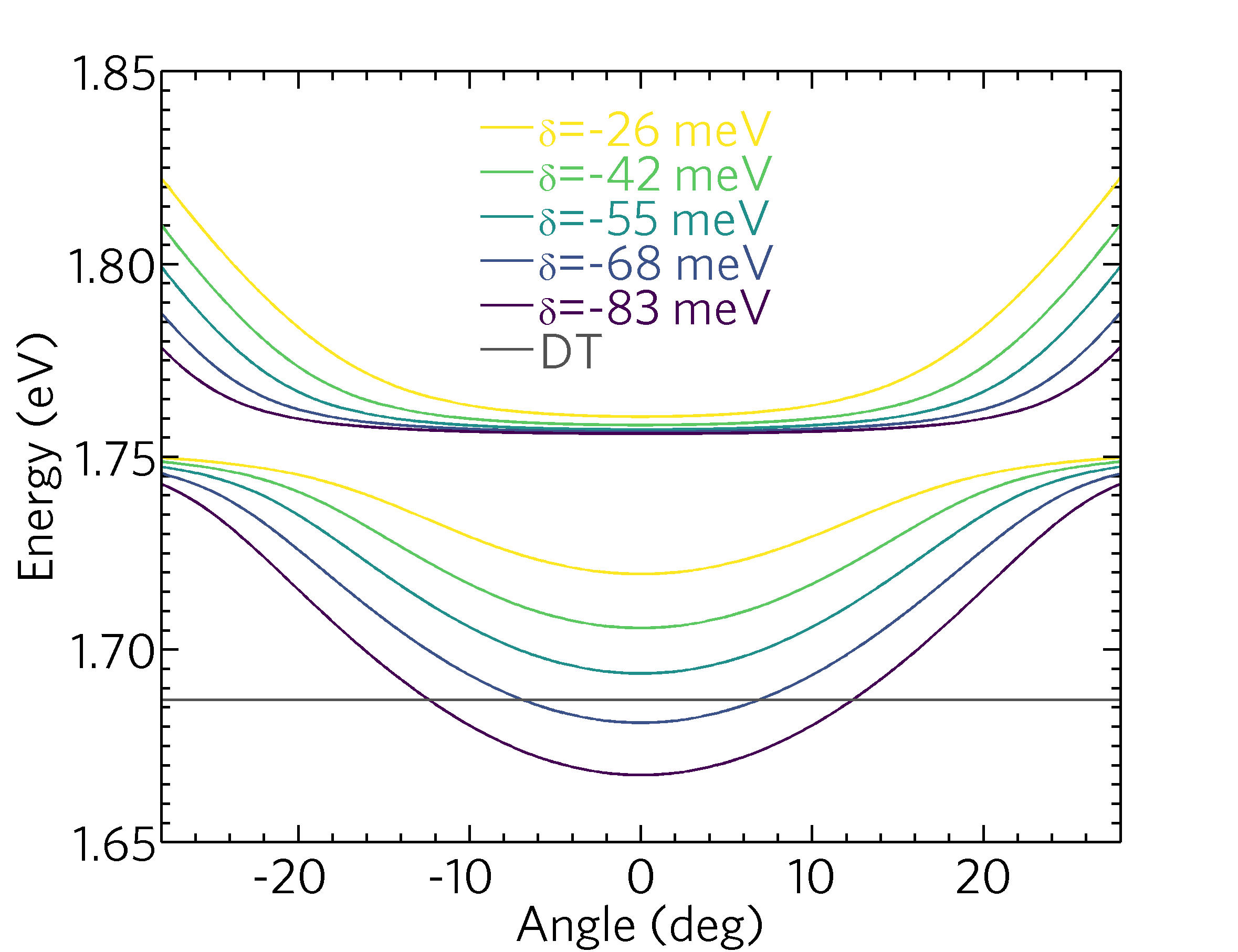}
\caption{Exciton-polariton dispersion relations for different detunings determined on the basis of  reflectance measurements with a marked energy of the dark trion state, $DT$.}
\label{im:Fig1SI}
\end{figure}

\begin{figure}[h]
\centering
\includegraphics[width=.41\textwidth]{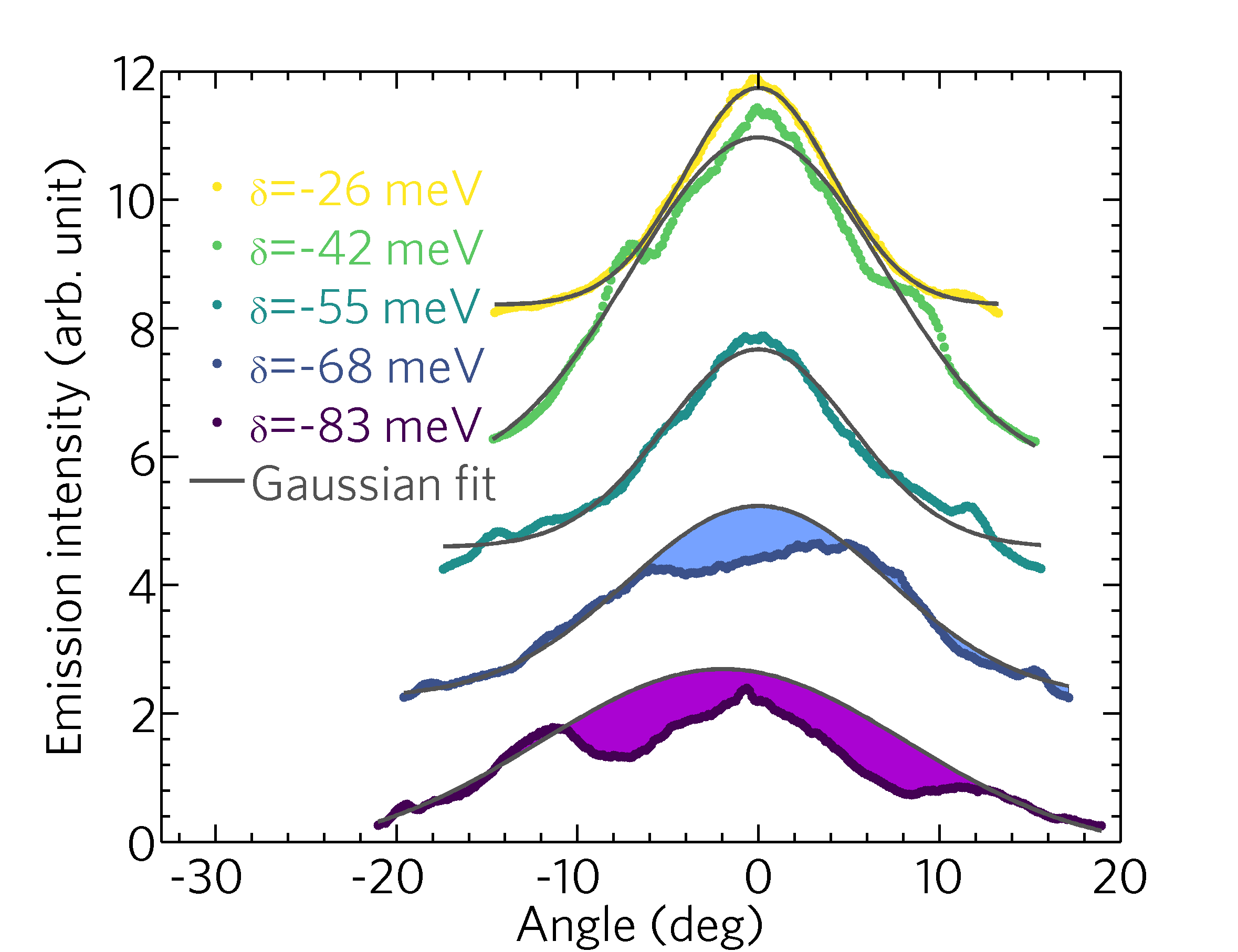}

\caption{Sum of the emitted light intensities detected in both circular polarizations for $\sigma^+$ excitation with Gaussian fits.}
\label{im:Fig2SI}
\end{figure}

A set of calculated dispersion curves corresponding to different exciton--photon detunings discussed in the main text is shown in Fig.~\ref{im:Fig1SI}. As can be seen, the lower polariton branches  for three most positive investigated exciton--photon detunings lay above the dark trion ($DT$) energy. Presented in Fig.\,\ref{im:Fig2SI}, the total emission intensity detected in both circular polarizations for $\sigma^+$ excitation displays in these three cases an  identical dependence on the observation angle. Maximum intensity is observed close to the zero emission angle, which corresponds to the lowest-energy polariton state at the bottom of the dispersion curve. With increasing the observation angle, the intensity decreases monotonically in the manner resembling a Gaussian bell curve. The situation differs for the two most negative detunings. The total intensity of emitted light for $-68$\,meV is nearly constant at low observation angles. Fitting a Gaussian function to the signal recorded at high emission angles shows a significant decrease in the emission intensity close to normal incidence. This angle range corresponds to the crossing between the energy of the lower polariton branch and the dark trion state at $1.687$\,eV in Fig.~\ref{im:Fig1SI}. An analogical emission suppression can be observed for the most negative detuning equal to $-83$\,meV. In this case, the expected crossing of the lower polariton's energy with the $DT$ state occurs for around 12\,deg and indeed, in the measured emission's intensity profile two symmetric dips at around 8\,deg and $-$8\,deg are clearly visible.

\newpage

\begin{figure}[h]
\centering
\includegraphics[width=.41\textwidth]{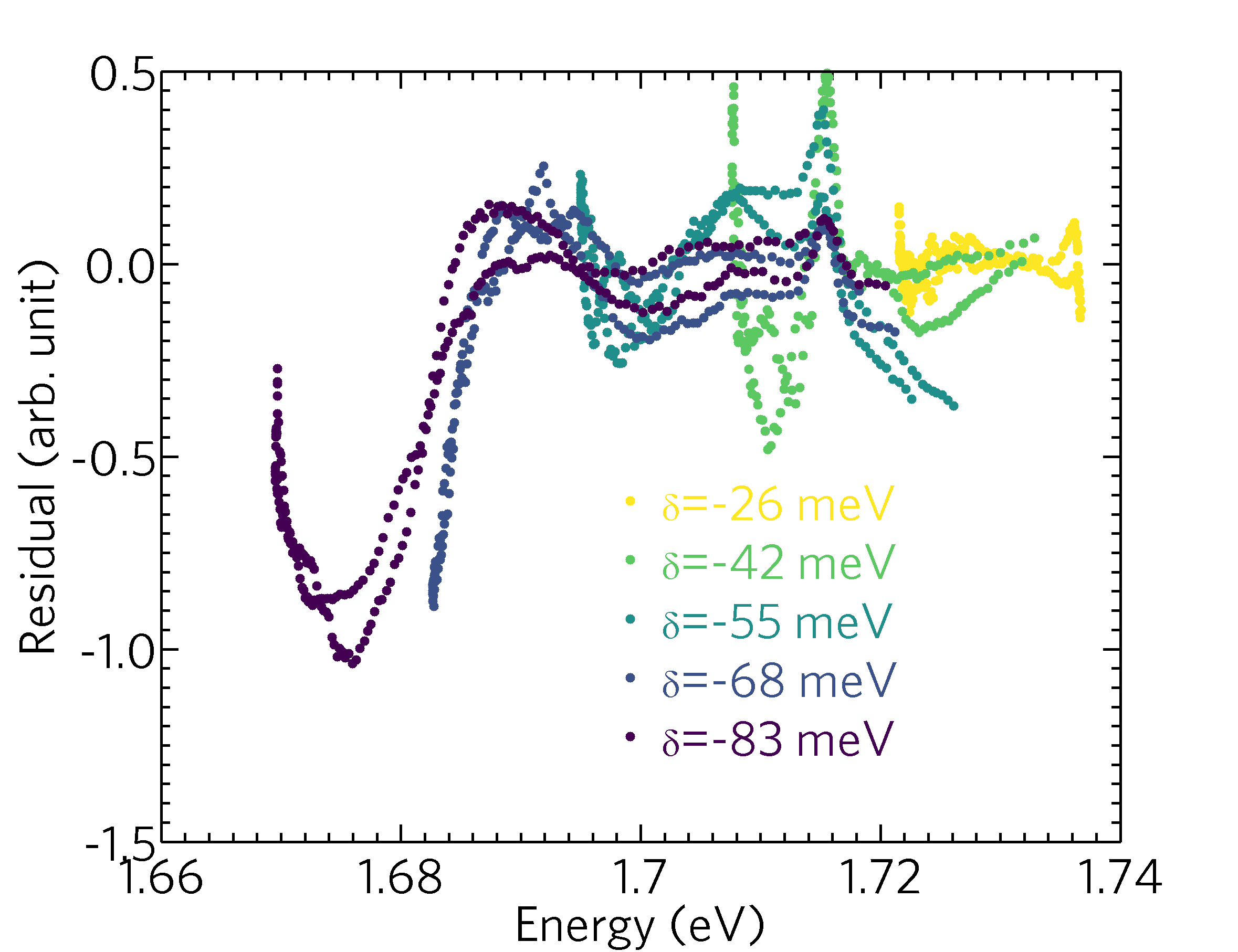}

\caption{Difference between the  measured emission intensity and the Gaussian fits presented in Fig.~\ref{im:Fig2SI}.}
\label{im:Fig3SI}
\end{figure}

To summarize the observed emission intensity's behaviour, a difference between expectations based on Gaussian function fits and experimental dependences is plotted as a function of energy in Fig.~\ref{im:Fig3SI}. For energies above 1.7\,eV it is close to zero. However, in the vicinity of 1.68\,eV a significant discrepancy appears, that is the observed emission intensity falls much below the expectation values, which we directly attribute to the photoluminescence suppression caused by resonant interactions of the exciton-polariton with the dark trion state.

